\documentstyle[psfig,preprint,aps]{revtex}

\begin{document}

\def\kk{{\bf k_1}}
\def\kg{{\bf k_2}}
\def\qq{{\bf q}}
\def\pom{{I\!\!P}}

% A useful Journal macro
\def\Journal#1#2#3#4{{#1} {\bf #2}, #3 (#4)}

% Some useful journal names
\def\NCA{\em Nuovo Cimento}
\def\NIM{\em Nucl. Instrum. Methods}
\def\NIMA{{\em Nucl. Instrum. Methods} A}
\def\NPB{{\em Nucl. Phys.} B}
\def\PLB{{\em Phys. Lett.}  B}
\def\PRL{\em Phys. Rev. Lett.}
\def\PRD{{\em Phys. Rev.} D}
\def\PRC{{\em Phys. Rev.} C}
\def\ZPC{{\em Z. Phys.} C}

% Some other macros used in the sample text
\def\st{\scriptstyle}
\def\sst{\scriptscriptstyle}
\def\mco{\multicolumn}
\def\epp{\epsilon^{\prime}}
\def\vep{\varepsilon}
\def\ra{\rightarrow}
\def\ppg{\pi^+\pi^-\gamma}
\def\vp{{\bf p}}
\def\ko{K^0}
\def\kb{\bar{K^0}}
\def\al{\alpha}
\def\ab{\bar{\alpha}}
\def\be{\begin{equation}}
\def\ee{\end{equation}}
\def\bea{\begin{eqnarray}}
\def\eea{\end{eqnarray}}
\def\CPbar{\hbox{{\rm CP}\hskip-1.80em{/}}}

\newcommand\noi{\noindent} 
\newcommand\la{\langle}
\newcommand\eps\varepsilon

\draft
\title{Unitarity Corrections to the Drell-Yan Process in the Target Rest Frame}
\author{M. A. Betemps
$^{\star}$\footnotetext{$^{\star}$E-mail:mandrebe@if.ufrgs.br},
 M. B. Gay Ducati
$^{\star \star}$\footnotetext{$^{\star \star}$E-mail:gay@if.ufrgs.br} and  M. V. T.
Machado $^{\star \star \star}$\footnotetext{$^{\star
\star \star}$E-mail:magnus@if.ufrgs.br}  }  
\address{ Instituto de F\'{\i}sica, Universidade
Federal do Rio Grande do Sul\\ Caixa Postal 15051, CEP 91501-970, Porto
Alegre, RS, BRAZIL}

\maketitle
\begin{abstract}
The unitarity corrections effects
encoded in the Glauber-Mueller approach are taken into account to
calculate the differential cross sections in the Drell-Yan process
in the rest frame. A detailed study of the Drell-Yan process in terms of
the $\gamma^*q$ transverse separation and the color
dipole size, and of the effective dipole cross section, is
performed and  compared with the available small $x$ data.
Estimates for the  Drell-Yan  cross section at RHIC energies are
presented and discussed.
\end{abstract}


\section{Introduction}

$ $

Both the  massive lepton pairs production in
hadronic collisions (Drell-Yan) and deep inelastic scattering (DIS)
at high energy are the most outstanding processes probing
the hadron structure. On the deep inelastic side, a large amount of 
work has been done to describe the copious data at medium and very small
Bjorken scaling variable $x$, based on perturbative QCD
\cite{CooperSarkar}. In the high energy domain it has been found
that important unitarity corrections should be taken into account
regarding the standard pQCD approach \cite{DGLAP}.   These phenomena
are
 currently denominated perturbative shadowing or saturation
effects \cite{mueller90,AGL,Victhesis,satmodels}. Concerning the Drell-Yan
sector, the pQCD tools have produced a reasonable theoretical
understanding of the main observables, despite the small data set
available at present \cite{DYreview}. The forthcoming
accelerator experiments (RHIC and LHC) will scan the high energy
limit of the hadronic reactions and open a new kinematic window,
i.e. smaller $x$ values. In particular, the quark-gluon plasma
(QGP), a new state of the hadronic matter predicted by QCD, is
expected to be found there \cite{Blaizot}. The theoretical
description of the QGP production is directly associated to a
complete knowledge on saturation effects and the transition
region to the high parton densities.  In a specific way, since the
production scheme for $J/\Psi$ is similar to the Drell-Yan one and
the latter does not contain final state effects, DY can be
considered as a baseline process to study  $J/\Psi$ suppression as a
signature of QGP formation \cite{Arcaferr}.  

In the fast proton system, the QCD
factorization theorem leads to describe the hadronic processes through the
convolution of the parton distribution functions (pdf's) with the
partonic subprocesses. The latter are completely calculated in pQCD
up to higher orders, whereas only the evolution in the factorization
scale  of the pdf's is determined.  Namely, the parton distributions
are solutions of the DGLAP evolution equations, whose formalism has
been successful in describing both DIS and DY data
\cite{CooperSarkar,DYreview}. Recently, an alternative way to study
electron-proton and hadronic reactions is claimed by the color dipole
picture considering the rest frame description based on
$k_{T}$-factorization \cite{mueller90,DYdipole1,DYdipole2}. Thus, the basic
blocks are the dipole light-cone  wavefunction and the dipole-target
cross section. Such an approach has produced an unified way to study
the mentioned processes, however its complete connection with the
standard DGLAP formalism is not provided yet and deserves further
studies.       

In the infinite momentum frame, the DY process
corresponds to the annihilation of  a quark (antiquark) from the projectile
with antiquark (quark) of the target into a virtual photon (vector
boson), which afterwards decays into a lepton pair
\cite{originalDY}. In the leading order (LO) calculation, the DY process 
has a simple electromagnetic character and it  can be promptly given
by QED theory. However, the perturbative QCD results at higher
orders modify this simple picture.  At present, pQCD
calculations have been developed up to the second order of the
strong coupling constant $\alpha_{s}$ \cite{vanNerwen}. For practical
considerations, in general the  involved next orders contributions
are taken into account by a phenomenological parameter, namely a
$K$ factor which is dependent on  the DY kinematic variables.   

In the rest frame, the DY process looks like a
bremsstrahlung of a virtual photon decaying into a lepton pair, rather
than a parton annihilation \cite{DYdipole2}.  The bremsstrahlung of
the virtual photon can occur after or before the interaction with the gluonic
field of the target. The  advantage of this formalism is that the
corresponding  cross section can be considered in terms of the same dipole
cross section extracted from small-$x$ DIS in the color dipole picture
\cite{Amirim}. At high energy, the unitarity corrections 
should be included in the dipole cross
section. Such effects have been considered, for example, in the 
phenomenological model of G.Biernat-Wusthoff (GBW) \cite{GBW}, which describes
DIS and $ep$ diffractive process with good agreement. We notice, however,
that the unitarity corrections to the inclusive observables, i.e. 
total cross section or $F_2$, can be hidden into the parametrization based 
on DGLAP approach, absorbed in the initial conditions, thus providing an 
excellent data description as seen in the updated NLO QCD fits \cite{newpdfs}.
More exclusive observables should be useful to clarify this important aspect. 
After this short remark, we 
proceed our argumentation.  The main
disadvantage in GBW is that a dynamical explanation of the saturation 
phenomenon is lacking. On the other hand, the Glauber-Mueller approach 
provides a theoretical development concerning parton
saturation \cite{AGL}, constraining the pQCD description of the dipole cross
section. Here, we make use of this formalism to perform a
description of the DY process in the rest frame.

The goal of this work is to perform a study of DY at high
energies considering the color dipole picture, in a similar way of
recent works \cite{Kop1}.  Our contribution is based on the use of the  
dipole cross section calculated in perturbative QCD, through the 
Glauber-Mueller formula \cite{AGL}, which encodes the unitarity
effects (saturation) in the parton densities. This approach takes
into account the multiple Pomeron scattering hypothesis in an eikonal
way keeping the unitarity of the considered process.  A
comparison between the phenomenological GBW dipole cross section and
the theoretical Glauber-Mueller one is presented, verifying that the
two approaches have different behaviors at higher energies. This is
due to the dynamical dependence on the gluon distribution in the
Glauber-Mueller approach, whose Born term recovers the DGLAP kernel 
in double log approximation (DLA).
The nonperturbative region, i.e. large dipole sizes contributions, 
is addressed considering the freezing of the gluon distribution
under the initial perturbative evolution scale $Q_0^2$. Then, we
present DY calculations   in the rest frame of the target at
leading order in a $pp$ collision and perform a comparison with the
low $x$ DY differential cross section from the E772 Collaboration
\cite{E772}. We also  produce  estimates for  the cross section at
RHIC energies.   

The outline of this paper is the following. In
the next section we present a brief review of the DY process in the
dipole color picture, discussing the range of validity for  this
approach and showing  the role played by the $\gamma^* q$
wavefunction. In Sec. (3) we present high parton  density
effects calculated from the  Glauber-Mueller approach, and
confront them  with the phenomenological GBW model. We estimate
the contribution of the saturation effects for the dipole cross
section in  high energies (LHC and RHIC). In
Sec. (4), a  parameter-free prediction to the differential DY cross
section for the  available data at small-$x$ and estimates to RHIC
are performed. Finally, in the last section  the
results are discussed and we present our conclusions.      

\section{Drell-Yan in the Color Dipole Picture} 

$ $

Before the description of the Drell-Yan process in the rest 
frame, we would like to review the main kinematical variables and the
standard calculations in the laboratory system. This is important to
clarify the connection between them and to emphasize the
asymmetry projectile-target  in the rest frame picture.

In the laboratory system, the lepton pairs are  produced in the Drell-Yan
reaction where partons from the projectile (fast proton) interact with the
proton target \cite{originalDY}.  Looking at the
parton level, a quark-antiquark pair annihilates into a virtual photon in
leading order $q\bar{q}\, \rightarrow \gamma^* \,\rightarrow l^{+}l^{-}$.
The symmetry between target and projectile is very clear, namely we cannot
distinguish a quark coming from the proton target or from the incoming beam.
The momentum fraction carried by the quark from the projectile is labeled $x_1$
and from the target is $x_2$. The partonic subprocess above  is well
known from QED, and the hadroproduction cross section is obtained folding the
partonic cross section with the quark (antiquark) densities evaluated at the
invariant $M^2$, the squared lepton pair mass, chosen here as the 
factorization 
scale $\mu_{\rm{fac}}^2$. Their evolution in $M^2$ is given by the standard
DGLAP evolution equations.  Therefore, the DY differential cross section in
leading order is given by 
\begin{eqnarray} 
\frac{d^2 \sigma^{DY}}{dM^2\,dx_F}
= \frac{4\,\pi \,\alpha^2_{\rm{em}}}{9\, M^2\,
s}\,\,\frac{1}{(x_1+x_2)}\,\,\sum_{f}e^2_f\,\left[q_f(x_1,M^2)\bar{q}_f(x_2,M^2)
+ \bar{q}_f(x_1,M^2) q_f(x_2,M^2) \right]\,, 
\label{dybreit}
\end{eqnarray} where
$q[\bar{q}]_f(x,M^2)$ are the corresponding quark (antiquark) densities with
flavour $f$ and squared charge $e^2_f$. The center of mass energy squared is
$s$ and the  usual notation is  
\begin{eqnarray} x_F & = &
x_1-x_2,\\ \tau & = & x_1x_2=M^2/s\,.
\end{eqnarray}
The momentum fractions are rewritten as
\begin{eqnarray}
x_1 & = & \frac{1}{2} (\sqrt{x_{F}^{2}+4\tau}+x_{F}),\\
x_2 & = & \frac{1}{2} (\sqrt{x_{F}^{2}+4\tau}-x_{F})\,,
\label{vardy}
\end{eqnarray} where $x_F$ is the longitudinal momentum  fraction, labeled Feynman $x$.
Indeed, $x_F$, $M$ and  $s$ are the kinematic variables experimentally
measured, whereas the partonic variables, $x_1$ and $x_2$, are reconstructed
from them.

When we consider the target at rest, the DY process  looks
like a bremsstrahlung: the quark from the projectile radiates a 
photon, which carries a fraction $\alpha$ of the light-cone momentum
of the initial quark, later decaying into the lepton pair [ see
Fig. (1)]. The interaction with the target can occur before or after
the photon emission. Thus, although  diagrammatically no dipole
to be present, the interference among graphs results in a product of
two quark amplitudes in the DY cross section, testing the external
gluonic field at two different transverse positions \cite{DYdipole2}.  
Therefore, a
remarkable feature emerging is that the $\gamma^*q-N$ interaction can be
described by  the same dipole cross section as in DIS \cite{Amirim}.

In the ($\alpha,\,\, r_{\perp}$) mixed 
representation, the photoabsorbtion cross section in deep inelastic
is described by the convolution of the wavefunctions,
$\Psi_{\gamma^*}$, from the virtual photon and the interaction dipole
cross section, $\sigma_{q\bar{q}}$. The wavefunctions are considered
taking into account the first photon  Fock state   configuration,
namely a $q\bar{q}$ pair. The dipole cross section is modeled
phenomenologically based on a matching between the hard and soft
pieces, constrained  by the DIS available data. The transverse separation of the $q\bar{q}$ 
pair is $r_{\perp}$, and each quark (or antiquark) of the dipole carries a momentum 
fraction $\alpha$  (or $1-\alpha$), from the incoming photon. The small 
dipole size configurations can be described  through pQCD, whereas the large
size ones belong to the nonperturbative domain. Hence, one can write
the photoabsorption cross section  as a function of the scaling
variable $x$ and photon virtuality $Q^2$ in the quantum mechanics
form \cite{DYdipole1},
\begin{figure}[t] 
\centerline{\psfig{file=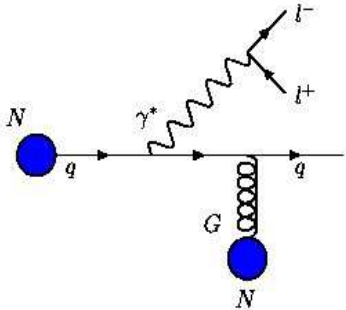,height=50mm,width=70mm}} 
\caption{~The Drell-Yan process in the rest frame, depicting one of the
possible interactions $\gamma^*q$-target (see text).}
\label{dybrem}
\end{figure}

\begin{eqnarray} \sigma_{T,L}(\gamma^*p
\rightarrow q\bar{q}) = \int \,d^2 r_{\perp} \,\int_0^1\,d\alpha
\,\,|\Psi^{T,L}_{q\bar{q}}(\alpha
,\,r_{\perp})|^2\,\,\sigma_{q\bar{q}}(x,r_{\perp})\,\,,
\label{dipdis} \end{eqnarray}
where $T,L$ indicate the transverse and longitudinal
contributions to the total cross section. In a similar way, the cross
section for radiation of a virtual photon from a quark after
scattering on a proton has the following factorized form in the color 
dipole picture
\cite{DYdipole2}   \begin{eqnarray}
\frac{d \, \sigma_{T,L}(qp\to q\gamma^* p)}{d\ln\alpha} =\int d^2 r_{\perp}\,\,
 |\Psi^{T,L}_{\gamma^* q}(\alpha,r_{\perp})|^2 \, \,   \sigma_{q\bar
q}(x_2,\alpha r_{\perp}), 
\label{gplctotal}
\end{eqnarray} where we have the same dipole cross section as in DIS. 
Here $r_{\perp}$ is the photon-quark transverse separation, $\alpha
r_{\perp}$ is the  $q\bar{q}$ separation and $\alpha$  is the fraction of 
the light-cone momentum of the initial quark taken away by the photon. 
We notice
the difference with the DIS case, where the dipole separation is just
$r_{\perp}$. Here, $\sigma_{q\bar{q}}$ is the cross section 
for scattering a $q\bar{q}$ pair  off a proton which depends on the $q\bar{q}$
transverse separation, and which should take into account the saturation
effects at high energy.

The physical interpretation of (\ref{gplctotal}) is similar to DIS in
the light-cone approach (LC). The projectile quark state is expanded in
its Fock space in the form \cite{DYdipole2}, 
\begin{eqnarray}
|q\rangle = Z_{2}|q\rangle + \Psi_{\gamma^{*}q}^{T,L} |q\gamma^{*}\rangle
+\dots  \label{fockspdy}
\end{eqnarray}
where here one has the  expansion in terms of the eigenstates from the 
quark projectile. Instead, in deep inelastic the  expansion is constructed from
the  eigenstates of the incident photon \cite{DYdipole1}. Here $Z_{2}$ is the 
renormalization constant.

The well known LC wavefunctions can be calculated in perturbation theory
\cite{DYdipole1,DYdipole2}, and depend on the transverse separations and
momentum fraction $\alpha$. They play an important role in the dilepton mass
$M$ dependence.  We take the same notation for the  LC wavefunctions from
\cite{Kop1},
\begin{eqnarray}
|\Psi_{\gamma^{*}q}^T(\alpha, r_{\perp})|^2 & = & 
\frac{\alpha_{\rm{em}}}{\pi^2}\left(m_{f}^{2}\alpha^{4}K_{0}^{2}(\eta
r_\perp)+[1+(1-\alpha)^2]\,\eta^2K_{1}^2(\eta r_\perp)\right)\,, \\
|\Psi_{\gamma^{*}q}^L(\alpha, r_{\perp})|^2 & = &
\frac{2\alpha_{\rm{em}}}{\pi^2}M^{2}(1-\alpha)^2K_{0}^2(\eta r_\perp)\,.
\label{LCpq}
\end{eqnarray}

The functions $K_{0}$ and $K_{1}$ are the modified Bessel functions and  the
auxiliary variable $\eta$, depending on the quark mass $m_f$, is given by
\begin{eqnarray} \eta^2 = (1-\alpha)M^2+\alpha^2 m_{f}^2\,.
\end{eqnarray}

The hadronic differential cross section for the Drell-Yan process is
expressed in a factorized form, embedding the partonic cross section, 
Eq. (\ref{gplctotal}), into the hadronic environment, in the 
following way \cite{DYdipole2},
\begin{eqnarray}
\frac{d\, \sigma^{DY}}{dM^2 \,dx_{F}}= 
\frac{\alpha_{\rm{em}}}{6\,\pi M^2}\,
\frac{x_1}{(x_{1} + x_{2})}\int_{x_1}^{1}\frac{d\alpha}{\alpha^2}
\sum_{i}e_{i}^{2}\left[q_{i}\left(\frac{x_1}{\alpha}\right)+\bar{q}_{i}\left(\frac{x_1}{\alpha}\right)\right]
\frac{d\sigma(qp \rightarrow q\gamma^*p)}{d\ln\alpha},
\label{dycsprimi}
\end{eqnarray}
where $e_{i}$ is the quark charge. In this frame we use standard kinematical variables $x_1=(2P_2.q)/s$ and 
$x_2=(2P_1.q)/s$, with $x_1x_2=(M^2+q_{T}^{2})/s$, where $P_1$, $P_2$ and $q$ are the four momenta of 
the beam, target and virtual photon, respectively. $M^2=q^2$ and $q_{T}^{2}$ are the dilepton invariant 
mass squared and the squared transverse momentum, respectively.

The frame dependence of the space-time interpretation of the DY process can be illustrated by different 
meanings of $x_1$ in different reference frames: we know that in the Breit frame, $x_1$ is the momentum 
fraction of the projectile quark (antiquark) annihilating with the target antiquark (quark). 
In contrast, evaluating the scalar product referred above in the target rest frame shows that the projectile 
quark carries momentum fraction $x=x_1/\alpha$ (which is larger than $x_1$), of the parent hadron, and 
correspondingly, $x_1$ is the momentum fraction of the proton carried by the photon. 
The variable $x_2$ is the momentum fraction of the proton carried by the gluon exchange in the $t$-channel.
We are benefited 
with the fact that the parton densities $q_{i}$ and $\bar{q}_{i}$ of the projectile enter in the 
combination $F_{2}^{p}$, which is the structure function of the proton. Therefore, we can rewrite the 
equation above in the following way,
\begin{eqnarray}
\frac{d\, \sigma^{DY}}{dM^2 \,dx_{F}}= 
\frac{\alpha_{\rm{em}}}{6\,\pi M^2}\,
\frac{1}{(x_{1} + x_{2})}\int_{x_1}^{1}\frac{d\alpha}{\alpha}
F_2^p\left(\frac{x_1}{\alpha}\right)
\frac{d\sigma(qp \rightarrow q\gamma^*p)}{d\ln\alpha},
\label{dycs}
\end{eqnarray}
where the summation of the longitudinal and transverse contribution 
was considered.
The factor $\alpha_{\rm{em}}/(6\,\pi M^{2})$ is due to the photon decay into
the lepton pair, coming from electrodynamics, the differential cross
section $d\sigma(qp \rightarrow q\gamma^*p)/d\ln\alpha$ is taken
from equation (\ref{gplctotal}) and our input to 
$\sigma_{q\bar{q}}$ in this work  \cite{AGL} is given by the
standard gluon distribution in the target corrected by saturation effects
in the high energy limit. In  Eq. (\ref{dycs}), the structure of the
projectile is described by the $F_2^p(x,Q^2)$ structure function. 

In the rest frame, the process is
asymmetric concerning projectile and target, in contrast with the symmetric 
picture in the Breit frame. The dipole color picture is valid
for small $x_2$ and it takes into account only the gluonic (sea quarks) sector 
from the target, disregarding its valence content. However, both valence 
and sea quarks in the projectile  are parametrized in the proton  structure
function in Eq. (\ref{dycs}) (for a complete discussion see Refs.
\cite{Kop1}). 
Although at present there is little range of experimental
measurements in the kinematical limit of validity of the color
dipole approach, it should provide reasonable
results when one considers smaller $x_2$ than the currently
available. The high energy accelerators LHC and RHIC will open a
wider kinematical window towards smaller $x_2$ values allowing
to test rest frame calculations properly.   

To conclude this
section, we analyse the behavior of the  wavefunctions in the
relevant kinematic variables. As it will be shown, they play the role
of a weight to the dipole cross section concerning the transverse
separations. In Eqs.
(\ref{dipdis}-\ref{gplctotal}), large $r_{\perp}$
configurations are suppressed in the integrated cross section,
controlling  the nonperturbative contributions (large transverse
distances domain) to the observables.  In the deep inelastic case, 
the LC wavefunctions dependence on the radius $r_{\perp}$ at fixed
photon virtuality $Q^2$ is discussed in Refs.
\cite{McDermott}. For the Drell-Yan case, the weight functions are
given by:  
\begin{eqnarray}   W_{\gamma_{*}q}^{T,L}(r_{\perp},
M^2)=r_{\perp} \int \frac{d\alpha}{\alpha}\,\,F_2^p
(x_1/\alpha,M^2)\,|\Psi_{\gamma^{*}q}^{T,L}(\alpha,r_{\perp})|^2 \,.
\label{weightf} \end{eqnarray} 

\begin{figure}[t] 
\centerline{\psfig
{file=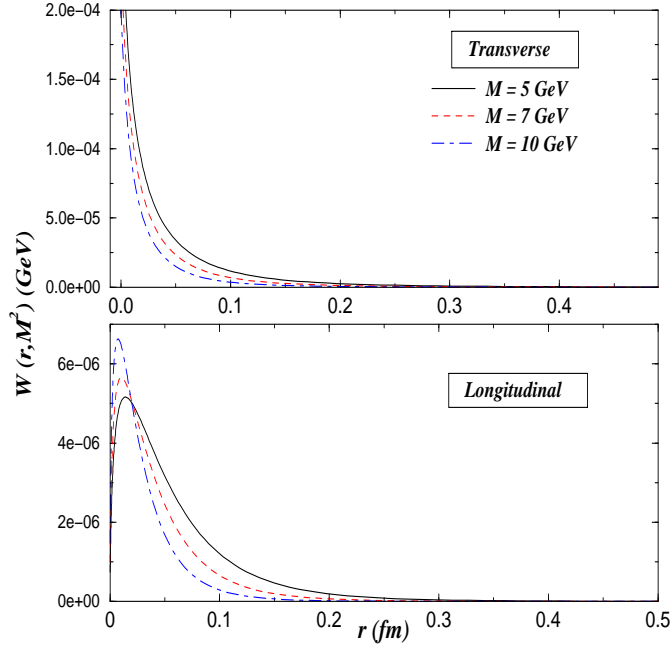,height=250.00pt,width=250.00pt}}
\caption{The longitudinal and transverse contributions for $W(r_{\perp},M^2)$
as a function of the $\gamma^*q$ transverse size $r_{\perp}$ at fixed
lepton pair mass $M$, for  $x_1 \approx x_F =0.525$.} \label{swvbe}
\end{figure}

In Fig. (\ref{swvbe}) one shows
separately the longitudinal and transverse results for
\,\,$W^{T,L}(r_{\perp},M^2)$ as a function of the photon-quark
transverse separation $r_{\perp}$ at fixed lepton pair mass $M$. The
choosen momentum fraction was $x_1 \approx x_F = 0.525$, since it is
a typical experimental value (see Sec. 4). Considering
this $x_1$ value, the proton structure function is insensitive
to the lepton pair mass range because it is in the scaling region. 
Regarding the quark mass, here one takes an effective light quark mass $m_f=0.2$ GeV in 
the wavefunctions.  

For the
transverse contribution, meaning the upper plot in Fig. (\ref{swvbe}), 
one verifies that the
weight function selects from small up to intermediate photon-quark sizes.
This means that it is selecting small dipole sizes ($\alpha r_{\perp}$)
in a similar way to deep inelastic scattering, since $x_1 \leq \alpha \leq 1$.
For our purpose here, the $x_1$ values reside close to $x_F$, then the 
conclusions in the 
following should hold when the weight factor is applied to the dipole 
cross section depending on 
$\alpha r_{\perp}$. A
steep increasing as $r_{\perp} \rightarrow 0$ comes from the
behavior of the function $K_1(\eta \,r_{\perp}) \sim 1/(\eta
\,r_{\perp})$  at this limit. Concerning the dependence on $M$, as
the invariant mass increases the contribution looks smaller. 

Regarding the longitudinal contribution, the lower plot in Fig. (\ref{swvbe}), 
the weight function
selects smaller dipole sizes (and $\gamma^*q$ transverse sizes) in
comparison with the transverse contribution. Moreover, the function is
narrower as $M$ increases, meaning that larger invariant mass scans
smaller $r_{\perp}$. A well known fact is that the longitudinal
contribution is higher twist, i.e. it is suppressed by a power of
$1/M^2$ when compared with the transverse one \cite{DYdipole1}. This
feature actually remains in the Drell-Yan case.  Moreover, the peaks
appearing in the plot are due to the balancing between the
asymptotic behavior at  $r_{\perp} \rightarrow 0$ of the function 
$K_0(\eta\,r_{\perp})\sim -\log(\eta\,r_{\perp})$ and the linear
$r_{\perp}$ factor in Eq. (\ref{weightf}).  

Having addressed the main features of the color dipole framework, namely
kinematic definitions and the description of Drell-Yan process in the rest
frame, in the next section one introduces our model for the dipole cross
section satisfying unitarity requirements.

\section{The Glauber-Mueller Approach}

$ $

The cross section for a color dipole-nucleon scattering is a well
known quantity, which was first proposed in the BFKL framework \cite{BFKL}. The
dipole interacts with the target through a perturbative Pomeron, described in
terms of the ladder diagrams. From the $k_T$-factorization framework
\cite{ktfact}, the scattering process can be written as the convolution of the
projectile impact factor and the unintegrated gluon structure function from
the target, whose dynamics is determined by the evolution kernel.  The
possible orderings in the transverse momentum $k_T$ in these graphs produce
the DGLAP or the BFKL dynamical evolutions. In particular, considering small
$r_{\perp}$ configurations from the dipole and the $k_T$-factorization one
obtains, 
\begin{eqnarray}  \sigma_{q\bar{q}}(x,r_{\perp}) = \frac{\pi^{2} \,
\alpha_s}{3}\,\, r_{\perp}^{2}\,xG_{N}^{DGLAP} (x,\frac{4}{r_{\perp}^{2}})
\label{dglapd} 
\end{eqnarray} 
where $xG_{N}^{DGLAP} (x,\tilde{Q}^2)$ is the
standard DGLAP gluon distribution at momentum fraction $x$ and virtuality scale
$\tilde{Q}^2=4/r_{\perp}^{2}$. An extensive phenomenology has been made
using the result above for the inclusive structure function and the vector
meson production \cite{McDermott}. In particular, we call attention to the
specific value of the scale $r_0^2$ appearing in the virtuality scale
$\tilde{Q}^2=r_0^2/r_{\perp}^{2}$. We use the $r_0^2 = 4$ throughout this
paper, however other values are equivalent at leading logarithmic
level \cite{McDermott}. 

A well defined feature from the data on $F_2$ and on the  gluon distribution at
high energies, i.e. smaller $x$, is that they present a steep increasing as
$x$ decreases. Indeed, experimentally $F_2 \sim xG \sim x^{-\lambda}$, where
the exponent ranges from 0.08 (Regge phenomenology) up to 0.5 (LO BFKL
calculations).  Such a behavior extrapolated to asymptotic energies violates
unitarity requirements and a control
should be considered. The scale where these effects start to be
important is associated to a region between hard and soft
dynamics \cite{DOLAnew} (pQCD versus Regge) or belonging to the high
density QCD domain \cite{AGL}. Here, we are interested in the last
case (for a recent review, see \cite{GayDucati}). 
In QCD, the taming of the gluon distribution at high energies is taken into
account through multiple Pomeron scattering encoded in the eikonal-like
frameworks \cite{Kovner}. Such a procedure provides the unitarization of the 
Born Pomeron cross section leading to a softer growth with energy. Indeed, the
asymptotic calculations have produced an unified $\ln(1/x)$ pattern for the
cross section and gluon function instead of a truly saturated one
\cite{asymptotic}.

In this work we use Eq.(\ref{dglapd}), where the standard DGLAP gluon
distribution is replaced by the modified one.  In the following we shortly 
review how the unitarity corrections are implemented through the 
Glauber-Mueller approach \cite{AGL}. The starting
point for the derivation is the  interaction of a virtual probe  particle, in
our case a virtual gluon,  with the nucleon. In the space-time picture
of this process, the virtual probe decays in a gluon-gluon ($GG$) pair having
transverse separation $r_{\perp}$. In the high energy limit, $r_{\perp}$ is
considered frozen during the interaction for $x<<1/(2m_NR_N)$, where the
nucleon $N$ has mass $m_N$ and geometric transverse size $R_N$.

The absorption cross section of a virtual gluon
($G^{*}$) with virtuality $Q^{2}$ and Bjorken $x$ can be written
in the form,
\begin{eqnarray}
\sigma^{G^{*}}(x,Q^2)=\int_{0}^{1}dz\int\frac{d^{2}r_{\perp}}{\pi}\int\frac{d^{2}b}{\pi}\,|\Psi^{G^{*}}
(Q^{2},r_{\perp},x,z)|^{2}\,\sigma^{GG}_N(x,r_{\perp})\,,
\end{eqnarray}
where $z$ is the fraction of the energy carried by the gluon, $b$ is
the impact parameter variable and $\Psi^{G^{*}}$ is the wavefunction
for the transversally polarized gluon generating the pair. The cross section
of the interaction  $GG$ pair with the nucleon  $\sigma^{GG}_{N}$
depends on energy $x$ and transverse separation $r_{\perp}$. This description
is valid in leading $\ln (1/x)$ approximation, however in the  double log
approximation (DLA) of perturbative QCD  one  obtains \cite{AGL},
\begin{eqnarray}
\sigma^{GG}_{N}(x,r_{\perp})=\frac{3\pi^{2}\alpha_{s}}{4}r_{\perp}^{2}xG^{DGLAP}
(x,\frac{4}{r_{\perp}^{2}})\,. \end{eqnarray}

The unitarity constraint to the cross section above is expressed by the
eikonal-like Glauber (Mueller) formula, hence the  the gluon structure
function can be written as \cite{GayDucati},
 \begin{eqnarray}
xG(x,Q^{2})=\frac{4}{\pi^{2}}\int_{x}^{1}\frac{dx^{\prime}}{x^{\prime}}
\int_{\frac{4}{Q^{2}}}^{\infty}\frac{dr_{\perp}^{2}}{\pi
r_{\perp}^{4}} \int \frac{d^{2}b}{\pi}\,\, 2\, \{
{1-e^{-\frac{1}{2}\sigma^{GG}_{N}(x^{\prime},r^{2}_{\perp}/4)S(b)}}\}\,.
\label{glaufor}
\end{eqnarray}

The explicit integration  limits for the $z$ (rewritten through the variable 
$x^{\prime}$) and transverse separation come from the physical kinematic range
allowed in the process (for detailed discussions, see \cite{AGL}). The impact
parameter $b$ dependence, is parametrized in the profile function $S(b)$. It
contains information about the angular distribution of the scattering in the
nucleon case and  the nucleon  distribution inside the nucleus in the nuclear
case. 

The Born term, in the expansion of Eq. (\ref{glaufor})
respect to $\sigma^{GG}_{N}$, provides  the DGLAP evolution in  double
logarithmic approximation (DLA). The remaining terms in the series contribute
to the saturation effects to the Born term. For simple calculations,
the   profile function $S(b)$ is parametrized as a Gaussian distribution, 
$S(b)=\frac{1}{\pi R_A^2}e^{-\frac{b^2}{R_A^2}}$, 
where $R_A$ is the target size, which is a free parameter to be determined
from data. Then, putting all together and performing the integration over
impact parameter $b$ in Eq. (\ref{glaufor}), one  obtains
\begin{eqnarray}
xG(x,Q^{2})=\frac{2R_A^{2}}{\pi^{2}}\int_{x}^{1}\frac{dx^{\prime}}{x^{\prime}}
\int_{1/Q^{2}}^{1/Q_0^{2}}\frac{d r_{\perp}^{2}}{r_{\perp}^{4}}\,\left(
\gamma _E +\ln [ \kappa_{G}  (x^{\prime}
,r_{\perp}^{2})] + E_{1}[\kappa_{G}(x^{\prime},r_{\perp}^{2})] \right) \,,
\end{eqnarray} 
where $\gamma _E$ and $E_1(x)$ are the Euler constant and the
exponential integral, respectively. The packing factor 
$\kappa_G=(3\pi\alpha_s r_{\perp}^2/2R_A^2)\,xG^{\rm{DGLAP}}$, sets the
scale where saturation effects are starting. Namely, the saturation scale
$Q_s^2$ is defined through the expression $\kappa_G(x,Q^2_s)=1$.

Since the Glauber-Mueller approach is valid in DLA, for practical reasons in
Refs. \cite{AGL,Victhesis} a procedure was introduced to extend the formalism
to the full experimental kinematic range available.  The final result contains 
the full DGLAP kernel corrected by contributions calculated in DLA, 
\begin{eqnarray} \nonumber
xG_{N}^{GM}(x,Q^{2})& &=xG(x,Q^{2})\,[Eq.(\ref{glaufor})] \\  +&
&xG^{DGLAP}(x,Q^{2})-\frac{\alpha_{s}N_{c}}{\pi}\int_{x}^{1}
\frac{dx^{\prime}}{x^{\prime}} \int_{Q_{0}^{2}}^{Q^{2}}\frac{dQ^{\prime
2}}{Q^{\prime 2}} x^{\prime}G^{DGLAP}(x^{\prime},Q^{2})\,, 
\label{gluondisttot}
\end{eqnarray}
where this modification is necessary to obtain a realistic approach in the
region of not very small-$x$. The above equation includes
$xG^{DGLAP}(x,Q^{2})$ as the initial condition for the gluon
distribution and gives $xG^{DGLAP}(x,Q^{2})$ as the first term of the
expansion with respect to $\kappa_{G}$. One needs to subtract the Born term  of
Eq. (\ref{glaufor}) in order to avoid double counting, which is the meaning 
the last term in equation above.

From the discussions and definitions above, we should use as dipole cross
section in our further calculations the following expression,
\begin{eqnarray}
\sigma_{q\bar{q}}^{GM}(x,r_{\perp})=\frac{\pi^{2}
\alpha_s}{3}\,\,r^{2}\,xG_{N}^{GM} (x,\frac{4}{r^{2}_{\perp}})\,.
\label{dipGM}
\end{eqnarray}

The resulting corrected gluon distribution (Eq. (\ref{gluondisttot})) has 
been applied for a comprehensive phenomenology in DIS process, considering 
the formulae above  as the gluon input for the observables calculated in the 
Breit frame (structure functions, $F_2$ slope, etc.). Recently, the GM dipole 
cross section has been applied in calculations of the DIS structure 
functions in 
the dipole color picture (see,  Ref.  \cite{mbgmvtm}). For instance, the 
structure functions description, in particular 
the latest $F_2$ data, can be seen at figures (3) and (5) of the 
Ref. \cite{mbgmvtm}, using that formalism.

\begin{figure}[h] 
\centerline{\psfig{file=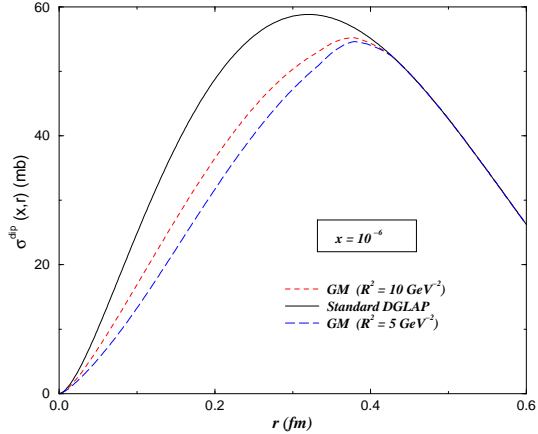,width=70mm,angle=-90.0}}  
\caption{The color dipole cross section as a function of the dipole size 
$r=\alpha r_{\perp}$ at fixed $x_2=10^{-6}$. The solid line corresponds to the 
result without unitarity corrections; the dotted line is the GM prediction using 
$R^2=10$ GeV$^{-2}$, and the dashed one is for $R^2=5$ 
GeV$^{-2}$.  } \label{gmdipcs} \end{figure}  

Now, one discusses in a detailed way the main characteristics
emerging from the Glauber-Mueller dipole cross section. To do
this, in Fig. (\ref{gmdipcs}) one shows the Glauber-Mueller dipole
cross section as a function of dipole transverse size $r=\alpha
r_{\perp}$ at fixed momentum fraction $x_2$. For sake of a better
illustration on the partonic saturation effects, one takes a very
small value for $x=10^{-6}$. Hereafter, one is using the GRV gluon
distribution at leading order \cite{GRV94}. 
Here we  use the GRV94 parametrization, since it has been
considered as a robust in a comprehensive phenomenology concerning 
unitarity corrections \cite{AGL,Victhesis,mbgmvtm,Vicslope,Gotsman}, 
which includes a 
related $R$ determination, intrinsic to the Glauber-Mueller approach 
\cite{Vicslope}.
The use of the others pdf's \cite{newpdfs}, implies in a determination 
of the corresponding  value of the  parameter $R$ as well as the enhancement 
of the already present uncertainty about the  nonperturbative contribution. 
For sake of illustration, in order to  test the sensitivity to 
the choice of pdf's set \cite{newpdfs}, in the Ref. \cite{McDermott}
such a study has been performed and it has been found that the deviations  
among the parametrizations are more important in the high virtuality region 
(very small $r$). However, in this region the color transparency behavior 
$\sim r^2$ dominates, absorbing possible sensibility in the specific selected parametrization.

\begin{figure}[ht] 
\centerline{\psfig{file=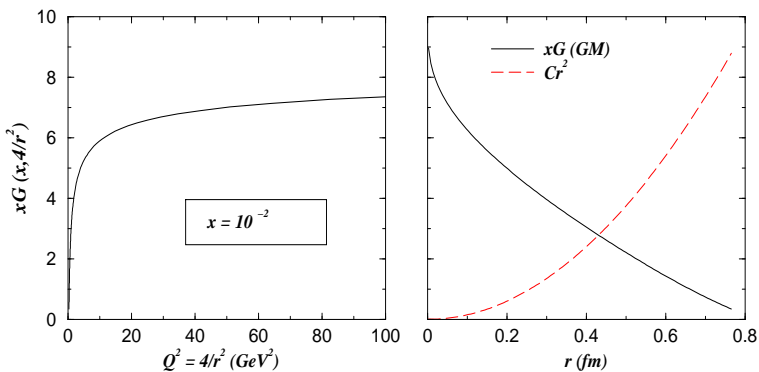,height=50mm,width=100mm,angle=-90.0}}  
\caption{The plot on the left shows the GM gluon distribution (GRV 
parametrization input) as a function of the scale $\tilde{Q^2}=4/r^2$
 at fixed $x=10^{-2}$. On the right, GM gluon distribution versus $r$
 and the color transparency behavior $\sigma_{dip} \sim C\,r^2$ (for 
illustration one uses a free normalization $C=15$). } \label{gmdetail}
\end{figure}

The solid line
corresponds  to the standard DGLAP calculation (without
saturation), Eq. (\ref{dglapd}), whereas the remaining ones result
from unitarity corrections for two different target sizes. The
general shape in terms of the dipole size comes from the balancing
between the color transparency $\sigma \sim  {r^{2}_{\perp}}$
behavior and the gluon distribution shape. These features are
depicted in the  plots in Fig. (\ref{gmdetail}), where one verifies
a visible scaling  of $xG(x,\tilde{Q}^2)$  versus $\tilde{Q}^2$
(left plot) and its dependence on $r=\alpha r_{\perp}$ (right
plot).

The difference in the strength of the unitarity corrections associated with the
target size is a well known fact. From the data analysis, its value ranges
from $R^2=5-10$ GeV$^{-2}$, where smaller radius produces strongest
corrections [ see Fig. (\ref{gmdipcs})]. For our calculations we
choose the low value $R^2=5$  GeV$^{-2}$, corroborated by studies in
the inclusive structure function and its derivative \cite{Vicslope}.

For sake of comparison, one considers the phenomenological model
of Ref. \cite{GBW} (GBW), which has produced a good description of
HERA data in both inclusive and diffractive processes. It is
constructed interpolating the color transparency behavior
$\sigma_{dip} \sim r^2_{\perp}$ for small dipole sizes and a flat
(saturated) behavior for large dipole sizes $\sigma_{dip} \sim
\sigma_0$ (confinement). The expression has the eikonal-like form, 
\begin{eqnarray}
\sigma_{q\bar{q}}(x,r)=\sigma_{0}\left[1-exp\left(\frac{r^{2}Q_{0}^{2}}
{4(x/x_{0})^\lambda}\right)\right], 
\end{eqnarray} where $Q^2_{0}=1$ GeV$^2$ and the three fitted parameters are
$\sigma_{0}=23.03$ mb, $x_{0}=3.04\,10^{-4}$ and $\lambda=0.288$ and 
$R_0(x)=(x/x_0)^{\lambda/2}$ is the saturation radius.
In GBW, saturation is characterized
by the $x$-dependent saturation radius $Q_s^2(x)=1/R^2_0(x)$ instead
of the scale coming from Glauber-Mueller,
$\kappa_G(x,Q_s^2)=1$, which can be easily extended for the nuclear
case \cite{AGL}. 

Although GBW and GM are distinct approaches, the small $x$ DIS data are 
equally well described by both models. In particular the structure functions
have been systematically described using the GM formalism, see for instance
Refs. \cite{Victhesis,mbgmvtm,Huanglusa,Gotsman}. The main advantageous feature
of GM in relation to the GBW is the dipole cross section providing a deep 
connection with the gluon distribution, the leading  quantity at high energies.
Concerning GBW, we point out the following shortcomings and disadvantages in 
comparison with GM approach:  $(a)$ it is strictly a parametrization 
available for the small 
$x$ HERA data; $(b)$ there is no direct connection with the gluon content; 
$(c)$ it does not match DGLAP evolution equations; 
$(d)$ it does not consider the impact parameter dependence of 
the process; 
$(e)$ it leads to a quite strong saturation scenario in contrast 
with the other available approaches; 
$(f)$ concerning the hadron-hadron collisions, using GBW to 
calculate the pion-proton total cross section (convoluting 
the dipole cross section with the pion 
wavefunction) it predicts non-realistic results, i.e. the 
cross section saturates at $\sim 23.03$ mb in high energies.

\begin{figure}[t] 
\centerline{\psfig{file=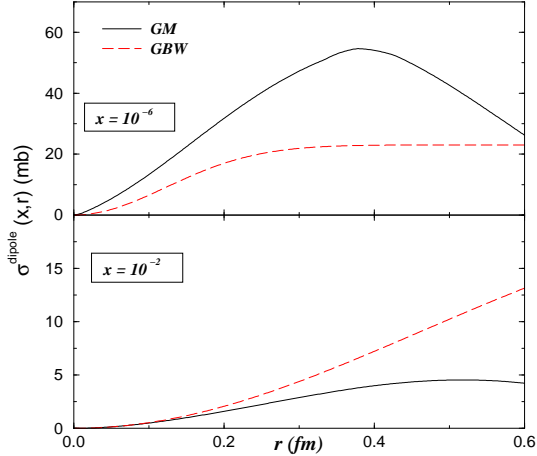,width=70mm}}  
\caption{The GM dipole cross section as a function of the dipole size 
$r=\alpha r_{\perp}$ at two typical $x_2$ values. The GBW result is also shown 
for sake of comparison.} \label{relcsgmwus} 
\end{figure}

In Fig. (\ref{relcsgmwus}) one shows the Glauber-Mueller dipole cross section
as a function of the dipole size $r=\alpha r_{\perp}$ for two typical $x_2$
values. In the lower plot, for $x_2=10^{-2}$, the GM cross section
underestimates the GBW one. However, as $x_2$ decreases the gluon 
distribution in the proton rises and the dipole cross section increases.
This feature is depicted in the upper plot, for a small $x_2=10^{-6}$, where
GM overestimates GBW by a factor two at intermediate $r \sim 0.3$. 
An immediate consequence
from the plots is that our prediction lies lower than GBW at
$x_2\approx 10^{-2}$ and higher for smaller $x_2$. We discuss these
features in a theoretical and phenomenological point of view when
performing the  comparison with available data in the next
section.   

\section{Results and Discussions}

$ $
This section is devoted to perform theoretical predictions for the available
data on DY process and the forthcoming ones from RHIC or LHC. In the previous
section, we presented a parameter-free Glauber-Mueller dipole cross section
which matches leading log gluon evolution and contains corrections from 
unitarity effects (parton saturation)  at higher energies. Therefore,
this provides a clear dynamical description of the
observables depending on the gluonic content of the target, also when it is a
nuclear one. 

Although perturbative QCD
provides reliable results at small distances (small dipole sizes), the
nonperturbative sector is still far from being completely
understood. The usual pdf's are
evolved from a perturbative initial scale $Q_0^2=M_0^2 \approx 1$
GeV$^2$, and there is little information about the behavior at $Q^2
\leq Q_0^2$, where the perturbative description is not even justifiable. In general one makes use of  Regge phenomenology to
estimate those contributions (see, for instance
\cite{McDermott}), and extrapolating to lower virtuality  regions
(large dipole sizes)  one needs an ansatz regarding the
nonperturbative sector.   

The use of the GRV94 parametrization \cite{GRV94} in our calculations, bearing
in mind that $Q^2=4/r^2$, meaning its evolution initial scale is $Q_0^2=0.4$
GeV$^2$, allows to scan dipole sizes up to
$r_{\rm{cut}}=\frac{2}{Q_0}$ GeV$^{-1}$ (= 0.62 fm). The cut off $r_{\rm (cut)}$ 
defines the transverse distance scale matching the perturbative and nonperturbative sector. For 
the most
recent parametrizations, where $Q_0^2 \sim 1$ GeV$^2$ ($r_{\rm{cut}}
\approx 0.4$ fm) the amount of nonperturbative contribution in the
calculations should increase. An additional advantage is that GRV94
does not include non-linear effects to the DGLAP evolution  since
the parametrization was obtained from rather large $x$ values. This
feature ensures that the parametrization does not include sensible 
unitarity corrections (perturbative shadowing) in the initial
scale.  

Now, we should introduce an ansatz for the large transverse
separation region. A more phenomenological way is to match the pQCD
dipole cross section with the typical hadronic one $\sigma_{\pi\,N}$
at $r_{\rm{cut}}$, for instance as performed in \cite{McDermott}.
Nevertheless, due to the large growth of the pQCD dipole cross section at
high energies and to take a more simple technical procedure
we choose an alternative way: the gluon distribution
is frozen at scale $r_{\rm{cut}}$, namely
$x\,G(x,\,\tilde{Q}^2_{\rm{cut}})$. Then, the large distance
contribution $r \geq r_{\rm{cut}}$ reads as,  
\begin{eqnarray}
\sigma_{q\bar{q}}^{GM}(x,r\geq r_{\rm{cut}})=\frac{\pi^{2}
\alpha_s}{3}\,\,r_{\rm{cut}}^{2}\,\,x\,G_{N}^{GM} (x,
\frac{4}{r^2_{\rm{cut}}})\,.  \label{ansatz}
\end{eqnarray} 
In a more rigorous analyses, one should substitute the freezening scale
$\tilde{Q}^2_{\rm{cut}}$ for the saturation scale $Q^2_s(x)$  to take into
account a realistic value of the gluon anomalous dimension in all kinematic
regions. 

Concerning our ansatz for the large $r$ region, one verifies in Ref. \cite{mbgmvtm} 
that it produces reasonable data description, mainly the normalization of the structure 
functions. An improvement for the non-perturbative contribution is performed 
there,
where a cut off in the $r$ integration ($0 \leq r \leq r_{cut}$) and  
the addition of a soft
Pomeron term are considered. However, if we do not introduce this improvement, the normalization 
of the structure functions remains
unaffected. For completeness, the consideration of a soft term for the non-perturbative 
contribution was also taken into account in Ref. \cite{McDermott}. However, for the same 
reasons above, we choose the technically more simple procedure.

\begin{figure}[t]
\centerline{\psfig{file=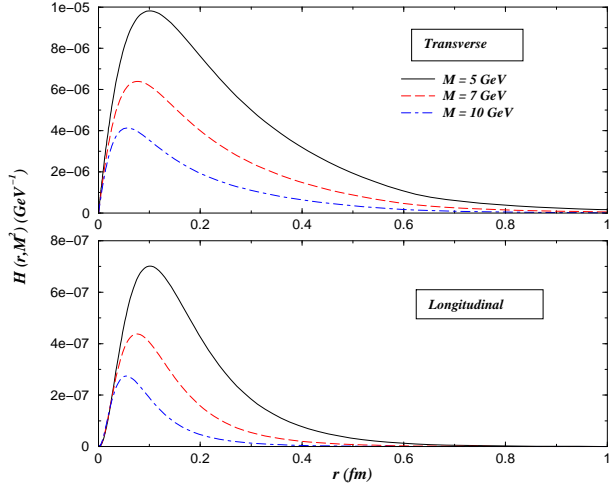,width=80mm,angle=-90.0}}
\caption{The profile $H_{T,L}(r_{\perp},M^2)$ as a function of the
$\gamma^* q$ transverse separation $r_{\perp}$ at typical mass $M$ values.
One uses $x_2=10^{-2}$ and GRV parametrization input.}
\label{rinteg} 
\end{figure}   

To illustrate the role played by the small and large transverse separations in
the description of the observables, in Fig. (\ref{rinteg}) one shows the
profile of the $r_{\perp}$-integration from Eq. (\ref{dycs}) as a function of
the $\gamma^* q$ transverse separation $r_{\perp}$ at typical mass $M$ values.
It is labeled here as $H_{T,L}(r_{\perp},M^2)$. The momentum transfer is fixed
at $x_2=10^{-2}$, since the low $x_2$ data available lie at this magnitude. For
the proton structure function $F_2^p(x_1/\alpha, M^2)$, describing the quark
content of the projectile, we use the ALLM updated parametrization
\cite{ALLM97} (good agreement with HERA data at large $x$). Both transverse and
longitudinal profiles are presented.

The main contribution for the profiles comes from the asymmetric peaks which
are shifted to larger $r_{\perp}$ as $M$ diminishes. For instance, in the
transverse profile, the peak lies at $r_{\perp} \approx 0.06$ fm for $M=10$ GeV
whereas at $M=5$ GeV it takes values $r_{\perp} \approx 0.1$ fm. As it was 
verified in Sec. (2), the longitudinal sector selects smaller transverse sizes
$r_{\perp}$ than the transverse one. Indeed, from the upper plot,  
non-zero contributions are obtained up to large  $r_{\perp} \approx 1$
fm. The higher twist character of the longitudinal piece is verified
through the magnitude scale of $H_L(r_{\perp},M^2)$. Due to the fact
that the nonperturbative sector dominates for $r_{\rm{cut}}= (\alpha
\, r_{\perp})_{\rm{cut}} \geq 0.62$ fm using the GRV input, a
significant part of the contributions comes from the soft region
where $\alpha$ is small (soft quark). This is in agreement with the
expectations that important soft contributions take place in
Drell-Yan process (see related discussions at \cite{Kop2} ).

\begin{figure}[ht]
\centerline{\psfig{file=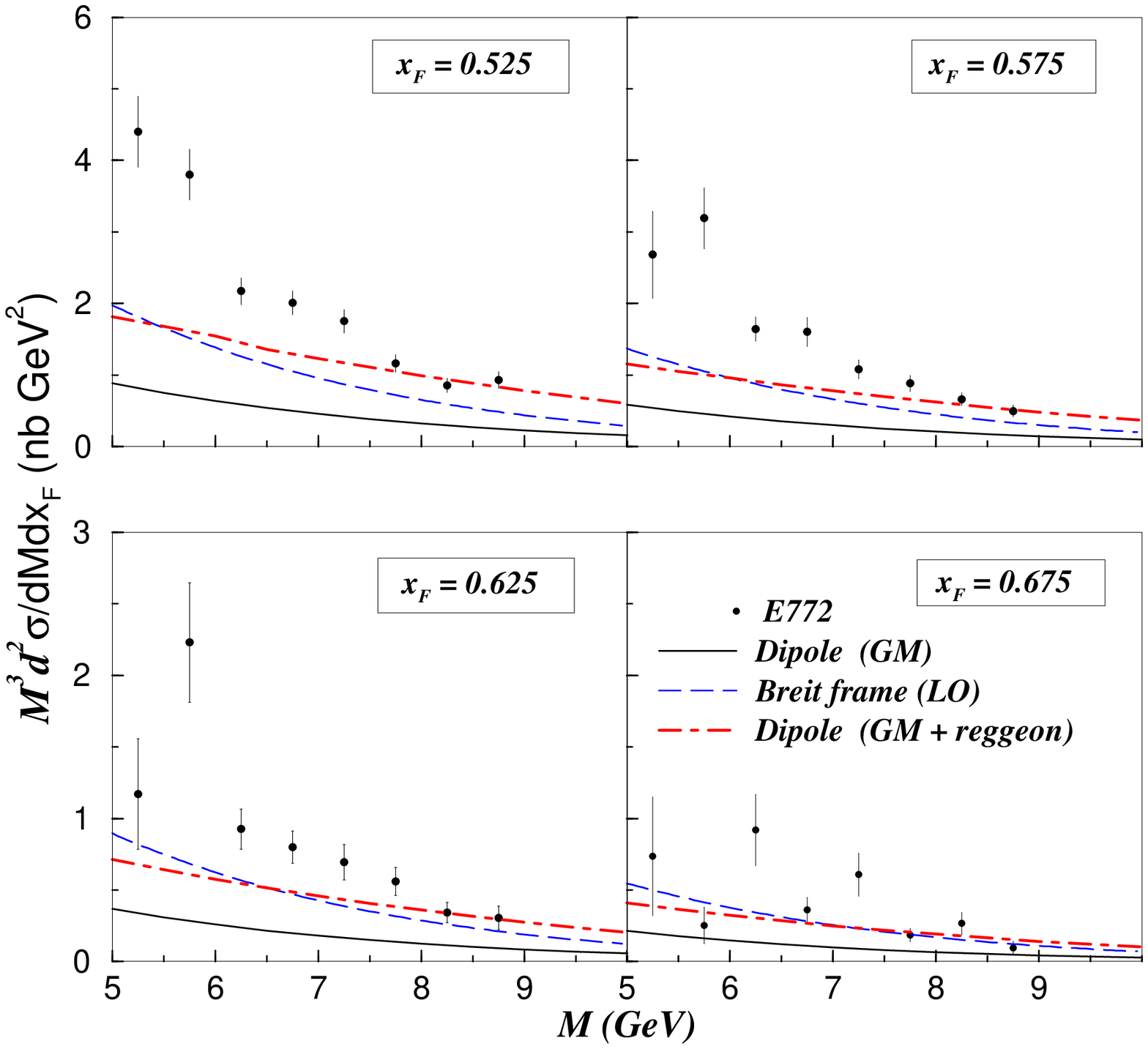,width=110mm}} 
\caption{ The DY differential cross section per nucleon  versus $M$ for the 
available small $x_2$ data [18] ($\sqrt{s}=38.8$ GeV) at fixed $x_F$ in
$pD$ reaction. The solid line corresponds to the Glauber-Mueller dipole cross
section. The dashed one is the LO Breit system calculation and the dot-dashed 
line corresponds to the Glauber-Mueller added of a reggeon contribution.} 
\label{diffdyE772} 
\end{figure}

Now, we are able to compare the results with the available data. Since the
color dipole picture is valid at small momentum fraction $x_2$, one needs
to select the experimental data covering this requirement. The lowest $x_2$
data were obtained in the fixed-target dimuon production at  E772
Collaboration \cite{E772}, where we select the points with $x_2<0.1$
following the similar procedure of \cite{Kop1}. In Fig. (\ref{diffdyE772}),
one presents the calculation Eq. (\ref{dycs}) using the
Glauber-Mueller dipole cross section (the solid line)  at fixed
$x_F$ and center of mass energy $\sqrt{s}=38.8$ GeV ($0.03 \leq x_2
\leq 0.09$). It should be stressed that this kinematical region
scans the validity limit of the color dipole picture. 
The curves underestimate similar calculations in Ref. \cite{Kop1}, which uses
the phenomenological GBW dipole cross section. Such a result is
actually expected from  our conclusions in the previous section where GM
underestimate GBW at $x_2=10^{-2}$ [see lower plot in 
Fig. (\ref{relcsgmwus})]. 

The experimental data analysed above are for $x>10^{-2}$. 
Therefore, the Eq. (\ref{dipGM}) is known to have corrections at larger $x$ 
values due to the exchange of quarks rather than gluons, in the $t$-channel, 
corresponding to a reggeon instead of a Pomeron exchange \cite{reggeon}. 
The secondary reggeon contribution corresponds to an amplitude with 
quark-antiquark pair $t$-channel exchange. The leading double-logarithmic 
asymptotics of such an amplitude was calculated in perturbative QCD in 
Ref. \cite{KirschnerLipatov}. The quark-antiquark cut occurs in the 
$j$-plane at $\omega_0(t)=\sqrt{2C_F \alpha_s/\pi}$, where 
$C_F=(N_c^2-1)/2N_c$ and $\alpha_s$ is the strong coupling constant. 
That value is very close to the phenomenological intercepts of the 
$\omega$, $\rho$ trajectories, i.e. $\alpha_R(0)\simeq 0.5$. Our expression for the dipole 
cross section, Eq. (\ref{dipGM}), considers only sea contribution for the process 
(gluon radiation), being equivalent in the Regge terminology to the hard Pomeron. The 
correspondent valence-like, which corresponds to the reggeon contribution, is lacking 
in our analysis above. In order to simulate the valence content in the calculations, we 
parametrize that piece in the following form \cite{reggparam}:
\begin{eqnarray}
\sigma_{I\!R}(x,r)=N_{I\!R}\,\, r^2 x^{0.4525}(1-x)^3\,,
\label{parametrizaR}
\end{eqnarray}
where we have used the reggeon intercept $\alpha_{I\!R}(0)=0.5475$ and the threshold 
factor for the large $x$ region \cite{reggparam}. To reproduce similar results as 
presented in Ref. \cite{Kop1}, one considers the constant value 
$N_{I\!R}=8$ (to obtain a $ \sigma_{I\!R}$ in GeV$^{-2}$). 
The $r^2$ factor  ensures the correct scaling.

In the plots (Fig. \ref{diffdyE772})  one shows also the LO
Breit system calculation, Eq. (\ref{dybreit}), which is the dashed line.
The color dipole result considering only the gluon content (sea quarks), 
Eq. (\ref{dipGM}),  lies below LO fast proton frame one at $x_2
\approx 10^{-2}$ (where the presented data are available). However, in this 
kinematical region the valence quark
content competes with the sea one and such a difference should be 
expected. Now, we introduce the valence content, parametrized in Eq. 
(\ref{parametrizaR}) and added to the  Eq. (\ref{dipGM}): as a result 
our data description  has been improved, equivalently to the 
calculations in \cite{Kop1} considering the GBW model.
 As $x_2$ decreases the gluonic content of the target
drives the observables and the color dipole considering only Eq. 
(\ref{dipGM})  should produce quite
reliable results.  We have verified this feature and have found 
that the reggeon contribution for the RHIC energies is completely negligible.

Here, some comments about higher order corrections are timely. The
color dipole approach results for total (virtual) photon cross
section are equivalent to those ones obtained by $k_T$-factorization
\cite{ktfact} in the leading logarithmic approximation. However, the
inclusion of higher order effects in the $k_T$-factorization
approach turns the equivalence incompatible: the conservation of
the transverse positions and sizes of the colliding objects is
violated \cite{Bialas}. Therefore, the introduction of  higher
orders contributions into the dipole cross section must be taken
with some care. Moreover, deep inelastic and Drell-Yan have a quite
different scenario concerning NLO and NNLO corrections. In DIS,
calculations considering up to NNLO resummations  have been
performed and it was found that they are small \cite{Altarelli}.
Instead, in Drell-Yan even the NLO calculations produce corrections
up to a factor of two, diminishing as the energy increases
\cite{DYreview,vanNerwen}. Keeping in mind the discussion 
above, at the moment we are unable to perform in the dipole color 
picture an equivalent NLO (Breit frame) calculation, 
since at the present the wavefunctions are not still available at NLO accuracy.

\begin{figure}[t]
\centerline{\psfig{file=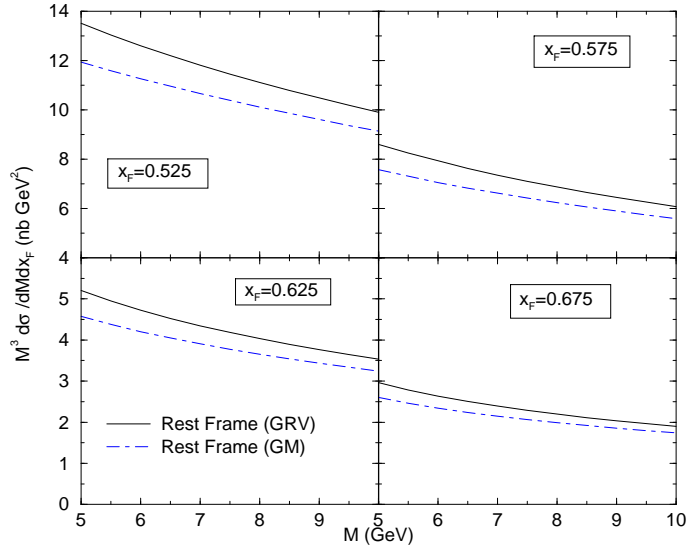,width=90mm}} 
\caption{The DY differential cross section per nucleon  versus $M$
for the RHIC energies ($\sqrt{s}=500$ GeV) at fixed $x_F$ in $pd$
reaction. The solid line corresponds to the Glauber-Mueller dipole
cross section whereas the dot-dashed one is the non-corrected DGLAP 
calculation.} \label{dyrhic} \end{figure}

In order to address the color dipole picture at high energies, 
the DY differential cross section
for RHIC energies, $\sqrt{s}=500$ GeV,  is shown in Fig.
(\ref{dyrhic}) for the same fixed $x_F$.  There, the $x_2$ reaches
values of order $10^{-4}$ and unitarity effects are important. The
solid lines are the Glauber-Mueller estimates, Eq. (\ref{dipGM}),  and the dot-dashed ones are
the rest frame calculations with DGLAP gluon distribution, Eq. (\ref{dglapd}).
The curves overestimate similar calculations in Ref. \cite{Kop1},
which uses the phenomenological GBW dipole cross section. Such a
behavior is expected from  our previous conclusions  where GM
overestimates GBW at smaller $x_2$ due to the growth of the gluon
distribution at higher energies  [see the upper plot in Fig.
(\ref{relcsgmwus})]. Concerning the rest frame non-corrected DGLAP input, the
Glauber-Mueller underestimates them due to the significant
corrections coming from unitarity effects (parton saturation).  Moreover, we have obtained a 
result almost similar to the LO Breit frame calculations at the RHIC energies, 
suggesting a good consistency in both frameworks. 
From the plots one
verifies that the deviations are more significant as $M$
diminishes, corresponding to smaller $x_2$ values. In absolute values, the corrections 
at RHIC energies  reach up to $\approx 20$ \%.  

As final comments,
we address additional advantages of the color dipole picture in
the DY case. For example, it allows to obtain the transverse momenta
$p_T$ distribution for the process already at leading order
calculation \cite{Kop1}. Instead, in the parton model the lepton
pair has no transverse momentum due to the assumption that in the
partonic subprocess the longitudinal momenta are bigger than the
transverse ones (partons are collinear). Therefore, an alternative way
to solve this trouble is to introduce an intrinsic $p_T$ for the
initial state interacting partons. However, such an assumption is not
sufficient to describe the measured $p_T$ distributions. Considering the 
Compton and the annihilation subprocess the leptons acquire  transverse momentum and the
$p_T$-dependence can be calculated in pQCD. The resummations produce
$\alpha_s\,\ln^2(m^2/p_T^2)$ terms which are large  as the transverse
momentum  goes to soft values $p_T \rightarrow 0$ (the perturbative
expansion breaks down). Thus, the color dipole description is a nice
tool to calculate that distributions since the above difficulties
are avoided.   One intends to address carefully this issue in a next calculation.

\section{Conclusions}
$ $

The Drell-Yan is an important process testing the quark (antiquark)
content of the hadron target. The measured observables are Lorentz invariant,
whereas the parton description is frame dependent. Calculations in the fast
proton system  have provided a perturbative understanding of DY up to higher
orders. On the other hand, the color dipole picture allows a simple
description of DY driven by the gluonic content (sea quarks) of the target.
The quark from the projectile radiates a photon decaying into a lepton
pair. The basic blocks in the color dipole are the LC wavefunctions and the
dipole cross section. The former is calculated from perturbation theory and
the latter one is modeled taking into account general properties of both hard
and soft domains.

We have found that the LC wavefunction for the $\gamma^*q$ configuration
plays the role of a weight function for the different transverse separations $r_{\perp}$
(as well as dipole sizes $\alpha r_{\perp}$) contribution to the process.
Namely, small transverse separations are selected by both the transverse and
longitudinal pieces. However, the transverse contribution can select non-negligible 
large sizes $r_{\perp}$. In addition, larger invariant mass $M$ scans smaller
$\gamma^*q$ separations.  Moreover, the longitudinal piece is higher twist
(suppressed by a factor $1/M^2$). These features are also present in the deep
inelastic case due to the similarity between the LC wavefunctions
expressions.   

Concerning the dipole cross section, here we consider the
Glauber-Mueller approach, which takes into account the corrections
to the standard DGLAP formalism due to the unitarity requirements.
The taming of the parton distributions (parton saturation) at high
energies is performed considering the multi-scattering assumption
from the Glauber-like (eikonal) formalism. We have found a distinct
behavior at both low and large $x_2$ when performing a comparison
with the phenomenological GBW model. The main source of the
deviation is that GM depends on the gluon distribution, which
increases as $x_2$ diminishes. These features produce distinct
results at current energies and in the forthcoming measurements. An
important verification is that a non-negligible amount of
non-perturbative contribution is present in the cross section.
Although the LC wavefunctions suppress large transverse separations,
a large cross section at small $x_{2}$ compensates the suppression
producing significant soft content. 

The current low $x_2$ available data lie in values ranging from $0.03 \leq x_2
\leq 0.09$, actually testing the validity limit of the color dipole picture.
Our results considering only GM dipole cross section underestimate the 
experimental measurements since color dipole
includes only the sea quark content (gluon radiation) from the target. In the
realistic case, for this kinematical region the valence and sea quarks have 
both a significant contribution in the cross section.  
We have parametrized the valence content through a reggeon exchange 
and the results turn out equivalent to the data description claimed 
in Ref. \cite{Kop1}. It was found also  that issues related to higher 
orders contributions  in the color dipole picture  should be taken carefully.

As the energy reached in the forthcoming experiments increases, 
the saturation effects should turn out to be more relevant. We 
perform estimates for the RHIC energies and
have found that the unitarity corrections are  important in 
the description of the cross section. We expect that such 
correction should be larger at LHC, since the $x_2$ values 
probed there would be smaller than in RHIC.

The quite simple scenario for DY process in the rest frame allows to extend the
approach to the nuclear case and also get information on $p_T$
distribution. The higher energies soon available will demand a well
established knowledge on the nuclear gluon distribution which can be
the input for the nuclear dipole cross section in the color dipole
framework. This approach should be a useful tool to perform pQCD estimates for
the future experimental measurements.

\section*{Acknowledgements}

M.A.B. and M.V.T.M. acknowledge useful discussions with Victor
Gon\c{c}alves.  M.V.T.M. also
acknowledges Martin McDermott (Liverpool University-UK) for useful
enlightenments. This work
was supported by CNPq, BRAZIL.


\begin{thebibliography}{99}

\bibitem{CooperSarkar} M. Klein, {\sl Int. J. Mod. Phys.} {\bf
A15S1}, 467 (2000). \\ A.M. Cooper-Sarkar, R.C.E. Devenish and A. De Roeck,
{\sl  Int. J. Mod. Phys} {\bf A13}, 3385 (1998).


\bibitem{DGLAP} Yu.L.
Dokshitzer. {\sl Sov. Phys. JETP} {\bf 46}, 641  (1977); \\ G. Altarelli and
G. Parisi. {\sl Nucl. Phys.} {\bf B126}, 298  (1977); \\ V.N. Gribov and L.N.
Lipatov. {\sl Sov. J. Nucl. Phys} {\bf 28}, 822  (1978).

\bibitem{mueller90}A.H. Mueller, {\sl Nucl. Phys.} {\bf 335}, 115 (1990).


\bibitem{AGL}A.L. Ayala, M.B. Gay Ducati, E.M. Levin, {\sl Nucl.  Phys.}
{\bf B493}, 305 (1997);  {\sl Nucl.  Phys.} 
{\bf B511}, 355 (1998).\\
M.B. Gay Ducati, V.P. Gon\c{c}alves, {\sl Nucl. Phys.} {\bf B557}, 296 (1999);
 {\sl Nucl. Phys.} (Proc. Suppl.) {\bf 79}, 302 (1999).


\bibitem{Victhesis} A.L. Ayala, M.B. Gay Ducati, E.M. Levin, {\sl Eur. Phys. J}
{\bf C8}, 115 (1999).
\\ A.L. Ayala, M.B. Gay Ducati, V.P. Gon\c{c}alves, Phys. Rev.
{\bf D59}, 054010 (1999).
\\ M.B. Gay Ducati, V.P. Gon\c{c}alves, {\sl Phys. Rev.} 
{\bf C60}, 058201 (1999); {\sl
Phys. Lett.} {\bf B466}, 375 (1999).

\bibitem{satmodels} L.V. Gribov,  E. M. Levin, M.G. Ryskin, {\sl
Phys. Rep.} {\bf 100}, 1 (1983);\\ A.H. Mueller,  {\sl Nucl.  Phys.} 
{\bf B558}, 285 (1999); A.H. Mueller, hep-ph/0111244. \\ 
Y. U. Kovchegov, {\sl Phys Rev.} {\bf D60}, 034008 (1999).\\
I. Balitsky, {\sl Nucl. Phys.} {\bf B463}, 99 (1996).\\
L. Mc Lerran and R. Venugopalan, {\sl Phys. Rev.} {\bf
D49}, 2233, 3352 (1994);
{\bf 50}, 2225 (1994); {\bf 53}, 458 (1996); \\
E. Iancu, A. Leonidov and L. Mc Lerran,  {\sl Phys. Lett.} {\bf B510},
133 (2001); Nucl.\ Phys.\ A {\bf 692}, 583 (2001);\\ E.~Ferreiro, E.~Iancu, 
A.~Leonidov
and L.~McLerran, [hep-ph/0109115].\\
J. Jalilian-Marian {\it et al.} {\sl Phys. Rev.} {\bf D59}, 034007
(1999);\\ N.~Armesto and M.~A.~Braun,
{\sl Eur. Phys. J.}  {\bf C20}, 517 (2001).



\bibitem{DYreview}  P.L. McGaughey, J.M. Moss, J.C. Peng,
{\sl Ann. Rev. Nucl. Part. Sci.} {\bf 49}, 217 (1999).

\bibitem{Blaizot} J.P. Blaizot, {\it Theory of the Quark-Gluon Plasma},
[hep-ph/0107131].

\bibitem{Arcaferr} N. Arnesto, A. Capella, E.G. Ferreiro, {\sl Phys.
Rev. } {\bf C59}, 395 (1999).

\bibitem{DYdipole1} N.N. Nikolaev, B.G. Zakharov, {\sl Z. Phys.} {\bf C49}, 607
(1991); {\sl Phys. Lett. } {\bf B260}, 414
(1991); {\sl Z. Phys.} {\bf C53}, 331 (1992).


\bibitem{DYdipole2} B.Z. Kopeliovich, In proceedings {\it Workshop
Hirschegg'95: Dynamical Properties of Hadrons in Nuclear Matter.}. Ed. by H.
Feldmeier and W. N\"orenberg, Darmstadt, p. 102 (1995). \\
S.J. Brodsky, A. Hebecker, E. Quack, {\sl Phys. Rev.} {\bf D55}, 2584 (1997).\\
B.Z. Kopeliovich,  {\sl Phys. Lett.} {\bf B447}, 308 (1999).

\bibitem{originalDY} S.D.~Drell, T.M.~Yan, {\sl Phys. Rev.
Lett. }{\bf 25}, 316 (1970). 

\bibitem{vanNerwen} A. Vogt, {\sl Phys. Lett.} {\bf B497}, 228 (2001).\\
P.J. Rijken, W.L. van Neerven, {\sl Phys. Rev.} {\bf D51},
44 (1995). 

\bibitem{Amirim} G.R. Kerley, M. McDermott,  {\sl J. Phys.} {\bf G26}, 683
(2000).\\ M.F. McDermott, Los Alamos preprint [hep-ph/0008260].

\bibitem{GBW} K. Golec-Biernat, M. W\"usthoff. {\sl Phys. Rev.} {\bf D59},
014017 (1999); {\sl Phys. Rev.} {\bf D60},
114023 (1999).


\bibitem{newpdfs} M. Gluck, E. Reya, A. Vogt,
{\sl Eur. Phys. J.} {\bf C5}, 461 (1998).\\
A. D. Martin et al.
{\sl Eur. Phys. J.} {\bf C}(to appear) (2002), [hep-ph/0110215].\\
H, L, Lai et al. 
{\sl Eur. Phys. J.} {\bf C12}, 375 (2000).


\bibitem{Kop1} B.Z. Kopeliovich, J. Raufeisen, A.V. Tarasov, {\sl Phys. Lett.}
 {\bf B503}, 91 (2001).

\bibitem{E772} E772 Collaboration, P.L.~McGaughey et al.,
{\sl Phys. Rev.} {\bf D50}, 3038 (1994); erratum {\sl
Phys. Rev.} {\bf D60}, 119903 (1999).

\bibitem{McDermott} M. McDermott et al., {\sl Eur. Phys. J.} {\bf C16}, 641
(2000).\\  L. Frankfurt, M. McDermott, M. Strikman, {\sl JHEP} {\bf 0103}, 045
(2001).

\bibitem{BFKL}
E.A. Kuraev, L.N. Lipatov and V.S. Fadin. {\sl Phys. Lett} {\bf B60} 50
(1975); {\sl Sov. Phys. JETP} {\bf 44} 443 (1976); {\sl Sov. Phys. JETP}
{\bf 45} 199 (1977);  \\  Ya. Balitsky and L.N. Lipatov. {\sl Sov. J. Nucl.
Phys. } {\bf 28} 822 (1978).

\bibitem{ktfact}  S. Catani, M. Ciafaloni, F. Hautmann, {\sl Nucl. Phys.} {\bf
B366}, 135 (1991).\\ J.C. Collins, R.K. Ellis, {\sl Nucl. Phys.} {\bf
B360}, 3 (1991).  

\bibitem{DOLAnew} A. Donnachie, P.V. Landshoff, {\sl Phys. Lett.} {\bf B437}, 408
(1998).

\bibitem{GayDucati} M.B. Gay Ducati. {\sl Braz. J. Phys.} {\bf 31}, 115 (2001) [hep-ph/0107116].

\bibitem{Kovner} A. Kovner, U. A. Wiedemann, {\sl Phys. Rev. } {\bf D64}, 114002 (2001).

\bibitem{asymptotic} A.L. Ayala, M.B. Gay Ducati, E.M. Levin, {\sl Phys. Lett.}
{\bf B388}, 188 (1996). \\ 
M.B. Gay Ducati, V.P. Gon\c{c}alves, {\sl Phys. Lett.} {\bf B502}, 92 (2001).

\bibitem{mbgmvtm} M.B. Gay Ducati, M.V.T. Machado, {\sl Phys. Rev.} {\bf D} (to appear), [hep-ph/0111093]. 

\bibitem{GRV94} M. Gluck, E. Reya, A. Vogt, {\sl Z. Phys.} {\bf C67}, 433
(1995).

\bibitem{Vicslope} M.B. Gay Ducati, V.P. Gon\c{c}alves, {\sl Phys. Lett.}
{\bf B487}, 110 (2000).

\bibitem{Huanglusa}Huang Z, Lu HJ, Sarcevic I, {\sl Nucl. Phys.} {\bf A637}, 79 (1998).

\bibitem{Gotsman} Gotsman et al., {\sl J. Phys. } {\bf G27}, 2297 (2001). 

\bibitem{ALLM97}H. Abramowicz, E. Levin, A. Levy, U Maor, {\sl Phys. Lett.} {\bf B269}, 465 (1991).\\ 
H. Abramowicz, A. Levy, [hep-ph/9712415].

\bibitem{reggeon} M.G. Ryskin, A.G. Shuvaev, [hep-ph/0203130] and references 
therein. 

\bibitem{KirschnerLipatov} R. Kirschner, L.N. Lipatov, {\sl Nucl. Phys.} {\bf B213}, 122 (1983); \\  R. Kirschner, {\sl Z. Phys.} {\bf C67}, 459 (1995).

\bibitem{reggparam}  A. Donnachie, P.V. Landshoff, {\sl Phys. Lett.} {\bf B518}, 63 (2001).

\bibitem{Kop2}  B.Z. Kopeliovich, {\sl Phys. Lett.} {\bf B447}, 308 (1999).

\bibitem{Bialas} A. Bialas, H. Navelet, R. Peschanski,  {\sl Nucl. Phys.}
{\bf B593}, 438 (2001); {\sl Nucl. Phys.} {\bf B603}, 218 (2001).

\bibitem{Altarelli} G. Altarelli, R.D. Ball, S. Forte, {\sl Nucl. Phys.}
{\bf B599}, 383 (2001).


\end{thebibliography}
\end{document}